\documentclass[10pt,conference]{IEEEtran}

\usepackage[utf8]{inputenc}
\usepackage[T1]{fontenc}
\usepackage{amsmath,amssymb,amsfonts}
\usepackage{graphicx}
\usepackage{textcomp}
\usepackage{xcolor}
\usepackage{booktabs}
\usepackage{array}
\usepackage{xurl}
\usepackage{microtype}
\usepackage{flushend}
\usepackage{csquotes}
\usepackage[hidelinks]{hyperref}

\usepackage[numbers,sort&compress]{natbib}
\makeatletter
\long\def\@makecaption#1#2{%
\ifx\@captype\@IEEEtablestring%
\bgroup\footnotesize\par\centering\@IEEEtabletopskipstrut{\normalfont\footnotesize #1}\par\egroup%
\setbox\@tempboxa\hbox{\normalfont\footnotesize\scshape #2}%
\ifdim \wd\@tempboxa >\hsize%
\bgroup\normalfont\footnotesize\scshape\noindent #2\par\egroup%
\else\hbox to\hsize{\normalfont\footnotesize\hfil\box\@tempboxa\hfil}\fi%
\addvspace{0.5\baselineskip}%
\@IEEEtablecaptionsepspace
\else
\@IEEEfigurecaptionsepspace
\setbox\@tempboxa\hbox{\normalfont\footnotesize {#1.}\nobreakspace\nobreakspace #2}%
\ifdim \wd\@tempboxa >\hsize%
\setbox\@tempboxa\hbox{\normalfont\footnotesize {#1.}\nobreakspace\nobreakspace}%
\parbox[t]{\hsize}{\normalfont\footnotesize\noindent\unhbox\@tempboxa#2}%
\else%
\hbox to\hsize{\normalfont\footnotesize\hfil\box\@tempboxa\hfil}%
\fi\fi}
\makeatother

\begin{document}

\title{Toward Evidence-Based Agentic~Software~Engineering}

\author{\IEEEauthorblockN{Sebastian Baltes}
\IEEEauthorblockA{\textit{Heidelberg University}\\
Heidelberg, Germany\\
sebastian.baltes@uni-heidelberg.de}
\and
\IEEEauthorblockN{Christoph Treude}
\IEEEauthorblockA{\textit{Singapore Management University}\\
Singapore\\
ctreude@smu.edu.sg}
}

\maketitle
\bstctlcite{IEEEexample:BSTcontrol}

\begin{abstract}
Task success rates and execution costs of agentic coding tools depend on both model and harness. The harness manages context, provides tools, and runs the agent loop. In empirical software engineering (SE) studies, comparisons that change model and harness together cannot isolate their effects. We qualitatively annotated 325 experiments reported in 146 peer-reviewed full papers published between 2025 and 2026. We examined the reported configurations and whether comparisons isolated harness effects, including harness configuration and version, model, tasks, and user interaction. The exact version of every harness used was reported for only 15 experiments. For five, the paper reported when the experiments were run, and for 31, authors shared execution traces for every session. Of the 226 experiments comparing harnesses, 35 never varied the harness in isolation. Based on this analysis, we derive requirements for a reference harness that the SE community can use for experiments. We call on the community to report all degrees of freedom of experiments with agentic coding tools and on SE venues to pilot dedicated tracks or special issues around a specified reference harness release. In these tracks, papers would hold the harness version fixed and vary features through separately versioned extensions to isolate feature effects and test their interactions.
\end{abstract}

\section{Introduction}
\label{sec:intro}
Task success rates and execution costs of agentic coding tools depend on both model and harness. An empirical software engineering (SE) study can attribute an outcome difference to the model or to the harness only if it identifies which model and which harness ran. The harness manages context, offers tools, enforces permissions, and runs the agent loop, in which a model chooses an action and receives its result. The \emph{inner harness} comprises the code, including the built-in system prompt and tool schemas, while the \emph{outer harness} is the configuration that users add, such as context files, settings, and hooks~\cite{lulla2026loop}. Model and harness interact as well. \citeauthor{DBLP:journals/corr/abs-2605-23950} ran three models on three harnesses and found that changing the harness reversed the ranking of two models in six of nine comparisons~\cite{DBLP:journals/corr/abs-2605-23950}. \citeauthor{DBLP:journals/corr/abs-2510-11977} found that \emph{Claude} models scored highest with one harness and OpenAI models with another~\cite{DBLP:journals/corr/abs-2510-11977}.

Existing reporting guidelines only partially cover the model and the harness together. For example, \citeauthor{DBLP:journals/corr/abs-2605-23950} propose a \emph{Harness Card} with seven layers, among them tools, the loop, traces, and the execution environment, but the card omits the harness version, the date of the experiments, and the cost~\cite{DBLP:journals/corr/abs-2605-23950}. The guidelines for studies involving large language models (LLMs) in SE ask for model versions, the date of the experiments, and execution traces~\cite{DBLP:journals/corr/abs-2508-15503}. \citeauthor{kapoor2024agents} demand that agent evaluations control for cost~\cite{kapoor2024agents}, and the \emph{ACM SIGSOFT Empirical Standards} add method-specific checklists~\cite{DBLP:journals/corr/abs-2010-03525}.
\citeauthor{DBLP:conf/icse/WilliamsKHAPS26} annotated which models and benchmarks the studies in 177 ICSE papers used~\cite{DBLP:conf/icse/WilliamsKHAPS26}, and \citeauthor{DBLP:conf/icse/SallouDP24} identified threats that LLMs pose to SE research~\cite{DBLP:conf/icse/SallouDP24}, but neither study reported the role of harnesses. The evaluation harnesses \emph{HAL}~\cite{DBLP:journals/corr/abs-2510-11977} and \emph{lm-eval}~\cite{biderman2024lessons} standardize benchmark execution, and \emph{Terminus 2}~\cite{DBLP:journals/corr/abs-2601-11868} and \emph{mini-SWE-agent}~\cite{minisweagent2026} hold a minimal harness fixed to compare models. Estimating harness effects requires the reverse, that is, comparisons that change the harness and hold the model and the tasks fixed.

For this paper, we qualitatively annotated 325 experiments involving agentic coding tools to assess what configurations the papers reported and which comparisons isolated harness effects (Section~\ref{sec:review}), organized by five degrees of freedom (Section~\ref{sec:model}). From our findings, we derive four requirements for a \emph{reference harness}, a pinned, extensible baseline shared by research groups across experiments so they can compare feature effects and interactions and build cumulative evidence. We assess which of 11 open-source harnesses meet these requirements (Section~\ref{sec:testbed}). We further suggest creating dedicated tracks or special issues around a specified reference harness release, so that teams can compare changes against the same baseline, and we plan an experiment on one of the harnesses in which five features are added as separately versioned extensions, to estimate feature effects and interactions while holding the harness fixed (Section~\ref{sec:plans}).

\section{Degrees of Freedom of an Experiment}
\label{sec:model}
To assess whether experimental comparisons of agentic coding tools isolate harness effects, we distinguish five \emph{degrees of freedom}: the outer harness configuration, the inner harness, the model, the tasks, and who answered agent requests during an agentic session. We define the outer harness using the configuration mechanisms documented by \citeauthor{DBLP:conf/aiware/GalsterMLATB26}~\cite{DBLP:conf/aiware/GalsterMLATB26}. Following the LLM guidelines~\cite{DBLP:journals/corr/abs-2508-15503}, we include model versions in the model degree of freedom and task set versions in the task degree of freedom. We added a fifth degree of freedom to capture who answered agent requests, after encountering studies with human or model responses.
Each \emph{condition} that a paper compares corresponds to a particular configuration of the five degrees of freedom. The outcomes for a condition come from \emph{sessions}, each covering the agent's work on one task from the first prompt until the task ends or a limit is reached.

\textbf{Outer harness} (\textsc{oh}). Exhaustively comparing outer harness configurations is impractical. Related work documents eight repository-level configuration mechanisms across five agentic coding tools, including context files, hooks, and Model Context Protocol (MCP) servers~\cite{DBLP:conf/aiware/GalsterMLATB26}. The documentation for \emph{Claude Code}, one of the most popular harnesses, listed 185 settings keys as of October 2026~\cite{anthropic2026settings}. If one were to treat these keys as independent binary choices, this would yield $2^{185}$ potential combinations for experiments.

\textbf{Inner harness} (\textsc{ih}). Harness updates can change outcomes even when the model is held fixed. \citeauthor{DBLP:journals/corr/abs-2607-03691} compared 35 \emph{Qwen Code} releases with the same model and found that tokens per task rose from about 391K to nearly 668K with no gain in task resolution rate~\cite{DBLP:journals/corr/abs-2607-03691}. The same study found 1.5 to 18 releases per week across five harnesses. Thus, a harness name can refer to different implementations even in a short period. Evaluating a component change alone does not establish how the harness will perform when other components change too. For example, \citeauthor{DBLP:journals/corr/abs-2604-25850} evaluated changes to tools, middleware, memory, and the system prompt separately and in combination. On hard tasks, the memory changes alone resolved more tasks than combining all four changes~\cite{DBLP:journals/corr/abs-2604-25850}.

\textbf{Model} (\textsc{mo}). A comparison of two conditions that differ in both model and harness cannot isolate a harness effect. We record the model that chooses actions and any additional models that the harness uses to produce or assess solutions. For each model, we include the version and the sampling and reasoning settings, such as temperature and reasoning effort.

\textbf{Tasks} (\textsc{ta}). Different task sets can produce different success rates even when the harness and model are held fixed. For example, \citeauthor{shayanfar2026} found that agents passed benchmark tests more often on tasks with reference solutions that changed files close together in the repository's directory tree~\cite{shayanfar2026}. Per experiment, we record the task sets and versions, the selected instances, and the task descriptions supplied to the agent. We include instructions that researchers added to the initial task prompt and any fixed follow-up prompts that researchers sent through a script during the session. We distinguish benchmark-provided execution environments from environments that researchers or the harness create. For example, a Docker image supplied with the benchmark is part of \textsc{ta}, while a container configured by researchers or the harness is part of the harness.

\textbf{User interaction} (\textsc{in}). Per experiment, we record whether a human, a model acting as the user, or no one answers agent requests during a session. We also record what task information the human or model receives and when they are allowed to answer an agent request.

\begin{figure}[t]
\centering
\includegraphics[width=\columnwidth]{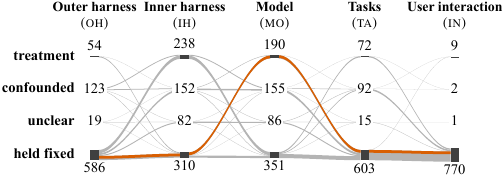}
\caption{Contrasts ($n=782$): Each path shows how the five degrees of freedom were coded in one contrast. Numbers and line widths count contrasts, and orange marks the most frequent combination of codes.}
\label{fig:model}
\end{figure}

\begin{table*}[t]
\caption{Reporting practices: \normalfont Column `Yes' ($n=325$) reports how many experiments answer each question. The percentage is the share among the experiments to which the question applies. A superscript refers to the question on which the marked question depends. Column `Disagreements' ($n=278$) reports how many experiments found by both annotations received different codes.}
\label{tab:review}
\centering
\footnotesize
\setlength{\tabcolsep}{2.7pt}
\renewcommand{\arraystretch}{0.94}
\begin{tabular}{@{}rlrr@{\hskip 7pt}rlrr@{}}
\toprule
\multicolumn{2}{@{}l}{\textbf{Disclosure question (\textsc{dq})}} & \textbf{Yes} & \textbf{Disagreements} & \multicolumn{2}{@{}l}{\textbf{Disclosure question (\textsc{dq})}} & \textbf{Yes} & \textbf{Disagreements} \\
\midrule
\multicolumn{4}{@{}l}{\itshape What did the study configure?} & 11 & How were modified or written tasks built?$^{10}$ & 207 (75\%) & 39 \\
1 & Can the outer harness be set up again? & 103 (32\%) & 49 & \multicolumn{4}{@{}l}{\itshape What can be checked afterwards?} \\
2 & Which tools could the model call? & 78 (24\%) & 33 & 12 & When were the experiments run? & 5 (2\%) & 0 \\
\multicolumn{4}{@{}l}{\itshape Which harness ran the tasks?} & 13 & Are traces shared for every task attempt? & 31 (10\%) & 1 \\
3 & Which version of each harness was used? & 15 (5\%) & 5 & \multicolumn{4}{@{}l}{\itshape Are the outcomes comparable?} \\
4 & Released, customized, or built? & 304 (94\%) & 8 & 14 & How was each measure computed? & 262 (81\%) & 48 \\
5 & Which parts were customized or built?$^{4}$ & 169 (77\%) & 23 & 15 & How much did repeated sessions vary? & 93 (29\%) & 20 \\
6 & Is the source code shared or open? & 163 (50\%) & 16 & 16 & What did a session cost, in which unit? & 165 (51\%) & 18 \\
\multicolumn{4}{@{}l}{\itshape Which model was used?} & 17 & Input and output tokens separately? & 28 (9\%) & 4 \\
7 & Is every model named with its version? & 238 (73\%) & 26 & 18 & Can a reader derive the money spent? & 61 (19\%) & 16 \\
8 & Which sampling and reasoning parameters? & 82 (25\%) & 25 & \multicolumn{4}{@{}l}{\itshape What else affects validity?} \\
\multicolumn{4}{@{}l}{\itshape Which tasks were run?} & 19 & Was contamination checked or mitigated? & 49 (15\%) & 24 \\
9 & Which task set ran, in which version? & 309 (95\%) & 10 & 20 & Was an open-weight model included? & 152 (47\%) & 7 \\
10 & Published, modified, or written tasks? & 317 (98\%) & 1 & 21 & Who answered agent requests? & 27 (8\%) & 62 \\
\bottomrule
\end{tabular}

\end{table*}

An \emph{experiment} is a set of conditions that a paper compares, possibly in the main text or across tables or figures. For example, \citeauthor{DBLP:conf/nips/BadertdinovGNSK25} compared 13 models using one harness in Table 1 of their paper~\cite{DBLP:conf/nips/BadertdinovGNSK25}. To compare performance across benchmarks, Table 2 of the same paper presents their \emph{SWE-rebench} results for seven of these models alongside their \emph{SWE-bench Verified} results. We count both tables as one experiment because Table 2 also extends the comparison of seven models to a second benchmark using the same harness.

Within an experiment, we count one \emph{contrast} per pattern of changed and unclear degrees of freedom, not per pairwise comparison. When several degrees of freedom change together, we count a separate contrast for each cause that the authors identify for the outcome difference. For example, all comparisons among 13 models on the same harness and tasks count as one contrast because only \textsc{mo} changes. Comparing task sets with the model held fixed counts as a separate contrast because \textsc{ta} changes. When both \textsc{ih} and \textsc{mo} change, a comparison that attributes an improvement to \textsc{ih} is counted separately from one that attributes the improvement to \textsc{mo}.

For each contrast, we use \emph{treatment} to denote the degree of freedom being tested. When only one degree of freedom changes, that change is the treatment. When several degrees of freedom change between the compared conditions, we assign the \emph{treatment} code to the degree of freedom that the authors report as causing the outcome difference. We code the other changed degrees of freedom as \emph{confounded}, because the comparison cannot separate the effects of those changes. For example, comparing \emph{Claude Code} with \emph{Claude Sonnet 4.5} against \emph{Codex CLI} with \emph{GPT-5} changes both harness and model. If the authors credit the harness, \textsc{ih} is the \emph{treatment} and \textsc{mo} is \emph{confounded}. When the compared conditions differ in several degrees of freedom and the paper does not attribute the outcome difference to a single degree of freedom, we code all changed degrees of freedom as \emph{confounded}.

Some authors attribute an outcome difference to their complete system without identifying any particular degree of freedom as its cause. We code the claimed cause as \emph{whole system}. For these contrasts, we assign \emph{treatment} to \textsc{ih} when harness implementations differ or \textsc{oh} when implementations match but configurations differ. Other changed degrees remain \emph{confounded}. This convention records the claim, not evidence that the harness effect was isolated.

We code a degree of freedom as \emph{held fixed} when the paper reports the same value for both conditions or describes no change, even if one or both values are unreported. This reflects the reported comparison, not verified equality. We use \emph{unclear} when it is uncertain whether a change applies to both conditions. For example, a paper may report raising the allowed tool-call limit for one condition without clarifying whether it also changed for the other. Comparing harnesses without identifying their models also makes \textsc{mo} \emph{unclear} because the models may differ.

We assess the reporting of experiment configurations separately from changes between conditions (Section~\ref{sec:review}). For example, \citeauthor{DBLP:conf/nips/BadertdinovGNSK25} used the same harness throughout their model comparisons but did not report a harness version~\cite{DBLP:conf/nips/BadertdinovGNSK25}. In this example, we code \textsc{ih} as \emph{held fixed} because the paper describes no change in the harness, although the exact release is unreported.

\section{Reporting of Agentic Coding Tool Studies}
\label{sec:review}
We qualitatively annotated experiments in peer-reviewed full papers published in SE, AI/ML, and related fields. On September 22, 2026, we searched DBLP titles and arXiv abstracts for terms related to software, harnesses, and agentic tools. Merging the results yielded 21,654 records. \emph{Claude Opus 5} and \emph{Sonnet 5} independently screened all titles and abstracts to identify empirical studies of agentic coding tools on SE tasks. We obtained 3,673 full texts for candidate papers selected by at least one screening model. Both models independently read the full texts and applied our inclusion and exclusion criteria to decide which papers to keep. \emph{Claude Fable 5.1} read the full texts of the 561 papers with conflicting decisions and settled whether to include each one. After further validation checks, removal of duplicate study versions, and venue and length filters, we retained 146 peer-reviewed full papers published in 2025 and 2026. These 127 conference papers and 19 journal papers report 325 experiments.
We then derived 21 disclosure questions (\textsc{dq}s) from the LLM guidelines~\cite{DBLP:journals/corr/abs-2508-15503}, adapting and extending them to assess how papers report agentic coding tool configurations and outcomes. We developed a codebook for identifying experiments and contrasts, coding the five degrees of freedom as described in Section~\ref{sec:model}, and answering the \textsc{dq}s.
The supplementary material provides the search queries, inclusion and exclusion criteria, selection counts, codebook, prompts, and annotations~\cite{supplement2026}.

We used \emph{Claude Code} (\emph{Opus 5.5}) subagents to annotate the paper full texts based on the codebook, a glossary, and a supplementary reference table documenting version formats for released harnesses encountered during the review. The subagents identified experiments and contrasts, recorded configurations and the authors' claimed causes, and answered the \textsc{dq}s with quotations. A script checked that quotations appeared on the cited pages and derived the codes for the five degrees of freedom deterministically from the subagents' contrast records. The first author resolved ambiguities, refined the codebook, and instructed the subagents to re-annotate papers affected by the rule changes. To assess coding reliability, we had \emph{GPT-6.1-Sol} repeat the annotation through API calls using the same codebook, without sharing the \emph{Opus} annotations.

To resolve disagreements about codes and experiment inclusion, we used separate \emph{Claude Code} (\emph{Opus 5.5}) subagents and \emph{GPT-6.1-Sol} API calls as judges. The judges read the full papers and codebook without being told which model had produced each annotation. Where both judges recommended the same code, we applied it. After the first author clarified disputed codebook rules, the judges reconsidered remaining disagreements in two further rounds using each other's reasoning. The first author decided to mark 32 unresolved codes as \emph{unclear} and include two experiments that the judges still disagreed about including. The final annotations contained 782 contrasts, grouped according to the rules in Section~\ref{sec:model}.

{\looseness=-1 Table~\ref{tab:review} summarizes the \textsc{dq} answers for these experiments and the annotation disagreements. Before disagreements were resolved, \emph{Opus 5.5} and \emph{GPT-6.1-Sol} identified 358 distinct experiments in total. For the 278 experiments identified by both models, we compared their 21 \textsc{dq} answers and codes for the five degrees of freedom. Disagreement was 8.5\% overall (95\% CI [7.5\%, 9.5\%], papers resampled) and 12.7\% for the codes assigned to the five degrees of freedom. Figure~\ref{fig:model} shows how the five degrees of freedom were coded in the 782 contrasts.

Exact versions of all harnesses run on the tasks were reported for 15 experiments (\textsc{dq}3). Five reported experiment dates (\textsc{dq}12), and 31 shared traces for every task attempt (\textsc{dq}13). In 238 experiments, the papers named the version of every model used (\textsc{dq}7). In 30 of these experiments, all reported model versions included a date. For example, \citeauthor{DBLP:conf/nips/BadertdinovGNSK25} reported the information requested by 14 of the 21 \textsc{dq}s for their main \emph{SWE-rebench} experiment but did not report a harness version or experiment date or share traces for every task attempt~\cite{DBLP:conf/nips/BadertdinovGNSK25}.\par}

{\looseness=-1 For \textsc{dq}3, we required a release number, tag, or commit identifier in the paper. We did not inspect linked repositories and archives. In the remaining 310 experiments, the papers did not give an exact version for one or more harnesses that the authors ran on the tasks. In 153 of these experiments, every harness was either open-source or had its source code linked in the paper (\textsc{dq}6).\par}

{\looseness=-1 Of the 226 experiments that compared harness implementations or configurations, 191 included at least one contrast that changed only the harness, whereas 35 did not. However, 141 included a contrast changing model or tasks together with the harness. In 115 experiments, at least one contrast changed the model together with the harness. In 50 of these 115 experiments, the authors attributed an outcome difference to their proposed agent as a whole, including its model and harness. Of those 50 experiments, 43 included a separate contrast that changed the harness alone. Among the 298 contrasts that changed more than one degree of freedom, authors credited their proposed system as a whole in 111 contrasts, without separating the effects of the harness, model, or other changes. In seven contrasts, the authors specifically credited the inner or outer harness. In 118 contrasts, the papers did not report a cause for the outcome difference.\par}

Our search could miss eligible papers with no arXiv record and no query term in their title. We did not search reference lists or citing papers. We excluded preprints without an eligible peer-reviewed full-paper version. We annotated only the paper full texts. Future work could include linked code and supplementary material to assess how much additional information about the experiments they provide. Screening, annotation, and judging relied mainly on \emph{Claude} models, so their errors may be correlated. Because models can disagree less with each other than with human annotators~\cite{DBLP:conf/msr/AhmedDTP25}, our 8.5\% disagreement rate does not establish how often the models would agree with human annotators. We developed and applied the codebook on the same papers, without testing it on an independent set.

\section{Requirements for a Reference Harness}
\label{sec:testbed}
A shared, versioned reference harness would help independent teams test features on the same code and build comparable evidence about their effects and interactions across studies. Based on our results, we propose four requirements:
\begin{enumerate}
\item \textbf{A pinnable release.} Researchers must be able to select an exact harness release and keep it fixed throughout an experiment. Recording that release lets other studies use the same implementation. Only 15 of the 325 experiments that our study identified reported exact versions for all harnesses that the authors ran on the tasks.
\item \textbf{Separately versioned extensions.} An extension needs its own version identifier so that researchers can change a feature while keeping the core harness release fixed. In 40 of the 782 contrasts, the inner and outer harness changed together and no other degree of freedom changed, so their individual effects could not be separated.
\item \textbf{Support for different models.} The harness must support both open-weight and commercial models. Researchers can compare conditions with and without a feature using one model, then repeat that comparison with another model. In 115 experiments that we annotated, the model changed together with the harness.
\item \textbf{Open-source.} Closed harnesses such as \emph{Claude Code} provide no source code for inspection or modification. A reference harness must therefore be open-source.
\end{enumerate}

We assessed 11 open-source coding harnesses against the four requirements listed above by inspecting documentation, release records, and relevant source files for the latest releases. %
We selected 10 harnesses that accepted arbitrary tasks described in natural language and were frequently used in the papers we analyzed: \emph{Aider}, \emph{Codex CLI}\textsuperscript{*}, \emph{Gemini CLI}, \emph{mini-SWE-agent}, \emph{OpenClaw}\textsuperscript{*}, \emph{opencode}\textsuperscript{*}, \emph{OpenHands}\textsuperscript{*}, \emph{Qwen Code}\textsuperscript{*}, \emph{SWE-agent}, and \emph{Terminus 2}.
We added the \emph{pi}\textsuperscript{*} harness~\cite{pi2026} because it is a popular open-source harness that focuses on minimality and extensibility. %
Superscript asterisks mark the six harnesses that satisfied all four requirements.
We selected \emph{pi} for the planned experiment (Section~\ref{sec:plans}) because its extensions can add native tools and replace the system prompt. For example, subagents and plan mode are provided as extensions rather than built-in features~\cite{pi2026}. This allows us to evaluate each feature by adding its extension to a baseline in which that feature is absent.

\section{Future Plans}
\label{sec:plans}
We plan an experiment using the \emph{pi} harness to estimate the effects and interactions of five separately versioned extensions: a permission gate, subagents, plan mode, a to-do list, and a compaction strategy. The first four add features absent from the inner harness, and the fifth replaces the built-in strategy. Prior studies varied one feature against a fixed harness, such as a codebase index~\cite{DBLP:journals/corr/abs-2606-22417}, a permission mechanism~\cite{bouras2026authority}, or harness rules~\cite{lewis2026same}. We will test 16 of the 32 combinations in a resolution~V fractional factorial design~\cite{montgomery2019doe}. This halves the number of configurations while estimating each extension's average effect and pairwise interactions, assuming that interactions involving three or more extensions are negligible. Interactions between pairs of features were the most common in a study of configurable systems~\cite{DBLP:conf/sigsoft/SiegmundGAK15}. We will run each selected combination with one open-weight and one commercial model at dated snapshots on the same tasks, using executable oracles and a fixed execution environment. Repeated task attempts will estimate success, cost, and feature interactions.
The experiment could be extended to \emph{opencode} and to built-in features in \emph{Claude Code} to test whether the findings generalize beyond \emph{pi}.

We suggest that the SE community should agree on a reference harness. Then, SE venues could issue calls for dedicated tracks or special issues on features implemented as extensions to that harness. Before issuing a call, organizers would select a harness release, publish a baseline configuration, and choose tasks from published SE benchmarks with executable oracles. Submitted studies would compare their extensions against the baseline on those tasks, holding the harness release and model fixed within each feature comparison. Authors would report all five degrees of freedom and the execution dates and share traces for every task attempt.
With a shared harness release and baseline, independent teams could repeat feature comparisons and combine evidence about their effects~\cite{DBLP:conf/icse/KitchenhamDJ04}.

\flushcolsend
\bibliographystyle{IEEEtranN}
\bibliography{references}

\end{document}